\documentclass[amsmath,amssymb,reprint]{revtex4-1}
\usepackage{graphicx}
\usepackage{caption}
\usepackage{subcaption}
\usepackage{dcolumn}
\usepackage{bm}
\usepackage[utf8]{inputenc}
\usepackage[T1]{fontenc}
\usepackage{mathptmx}
\usepackage{tikz}
\usetikzlibrary{shapes.callouts}
\tikzset{
    level/.style = {
        thick,
        blue,
    },
    connect/.style = {
        dashed,
        red
    },
    notice/.style = {
        draw,
        rectangle callout,
        callout relative pointer={#1}
    },
    label/.style = {
        text width=2cm
    },
    trans/.style = {
    thick, <->,shorten >=2pt,shorten <=2pt,>=stealth
    }
    }
\usetikzlibrary{arrows}
\usepackage{pgfplots}
\usepackage{braket}
\usepackage{soul}
\begin{document}
\title[]{Ultra-sensitive Single-beam Atom-optical Magnetometer using Weak Measurement Method}
\author{T.J. Vineeth Francis}
\affiliation{Tata Institute of Fundamental Research, Centre for Interdisciplinary Sciences,~Hyderabad 500107,~India}
\author{Rashmi Ranjan Suna}
\affiliation{School of Physics, University of Hyderabad, Hyderabad 500046,~India}
\author{P.K. Madhu}
\affiliation{Tata Institute of Fundamental Research, Centre for Interdisciplinary Sciences,~Hyderabad 500107,~India}
\author{Nirmal K. Viswanathan}
\email{nirmalsp@uohyd.ac.in}
\affiliation{School of Physics, University of Hyderabad, Hyderabad 500046,~India}
\author{G. Rajalakshmi}
\email{raji@tifrh.res.in}
\affiliation{Tata Institute of Fundamental Research, Centre for Interdisciplinary Sciences,~Hyderabad 500107,~India}

\date{\today}
\begin{abstract}
Ultra-sensitive measurement of the magneto-optical rotation, due to interaction of linearly-polarized light passing through room-temperature Rb$^{85}$ atoms, in response to change in longitudinal magnetic field ($\delta B_{z}$) is demonstrated using the weak measurement method. The polarization rotation angle measurement sensitivity ($\delta\phi$) of 16 $\mu rad$ and hence of the magnetometer of 1 nT, achieved using the weak measurement method is better than the balanced optical polarimetry results by a factor of three. The improvement in the measurement sensitivity is realized via optical amplification of the polarization rotation angle via spin-orbit coupled light beam-field. The method is devoid of external rf modulation, allows for optimal tunability of sensitivity depending on the dynamic range of the applied magnetic field and the sensitivity can be further enhanced by operating in the Spin Exchange Relaxation Free regime of alkali spin polarization.   
\end{abstract}
\maketitle
\section{\label{sec:level1}Introduction}
Optical magnetometry, the technique of using light beam to detect magnetic field and changes in it due to interaction in an atomic medium, is fast gaining significance in a variety of fundamental and applied research activities due to its unprecedented measurement sensitivities \cite{Budker2002,Budker2007,Taylor2008,Dang2010,Fan2018}. Atom-optical magnetometers working on the principle of resonant Faraday effect or Macaluso-Corbino Effect \cite{Macaluso1898,Macaluso1899}, have been known since late 1890s. For an on-resonance light beam propagating through the atomic medium, the magneto-optical rotation\cite{Budker2002} is a function of longitudinal magnetic field $B_{z}$ given by
\begin{align}
\phi \simeq \frac{\frac{2 g_{F} \mu_{B} B_{z}}{\hbar \Gamma}}{1 + \left(\frac{2 g_{F} \mu_{B} B_{z}}{\hbar \Gamma}\right)^{2}} \frac{l}{l_{o}}
\label{mo-rotation}
\end{align} 
where $g_{F}$ is Lande factor, $\mu_{B}$ is Bohr magneton, $\hbar$ is Planck's constant, $\Gamma$ is relaxation rate of excited state, $l$ is length of atomic medium and $l_{o}$ is absorption length. The sensitivity \cite{Budker2013}, $\delta B_{z}$, to change in longitudinal magnetic field  of the magnetometer is written as
\begin{align}
\delta B_{z} = \left( \frac{\partial \phi}{\partial  B_{z}}\right)^{-1} \delta \phi
\label{B-sensitivity}
\end{align}
where $\frac{\partial \phi}{\partial B_{z}}$ is the slope of magneto-optical rotation with respect to the longitudinal magnetic field $B_{z}$ and $\delta \phi$ is rotation sensitivity of the measurement, which is limited by the photon shot noise of the detector used. Methods to improve the magnetometer sensitivity includes enhancing the Faraday rotation of the atomic system to longitudinal magnetic field changes $\left(\frac{\partial \phi}{\partial B_{z}}\right)$ and ways to improve the detection sensitivity of the optical rotation signal $\left(\delta \phi\right)$. The former is achieved by tuning the longitudinal magnetic field at which maximum magneto-optical rotation occurs, given by
\begin{align}
B_{max} = \frac{\hbar \Gamma}{2 g_{F} \mu_{B}}
\label{B-max}
\end{align}
    Thus, for $B_{z}<<B_{max}$, the slope $\frac{\partial \phi}{\partial B_{z}}$ is inversely proportional to the relaxation rate $\Gamma$ of the atoms, making the sensitivity $\delta B_z$ proportional to $\Gamma$. In case of linear interaction of single beam of light with thermal atoms moving with  Maxwell-Boltzmann velocity distribution, $\Gamma$ is given by Doppler linewidth of the atomic transition which is about $600$~MHz at room temperature for Rb atoms, as measured from the absorption spectra. It is possible to achieve much smaller values of $\Gamma$ via non-linear processes such as formation of Bennett structures \cite{Bennett1962, Kimball2002} in the velocity distribution of the atomic system. Laser power above the saturation intensity can cause imbalance in the  distribution of angular-momentum state population due to optical pumping leading to hole burning effects, making $\Gamma$ comparable to the natural linewidth of absorption of $\approx 1-6$~MHz. Alternately, $\Gamma$ has been reduced by decreasing the relaxation rate of ground state atomic polarization via coherent effects \cite{Budker1998,Budker2000} in the atomic medium. Realization of this has led to the development of Spin Exchange Relaxation Free (SERF) Magnetometers \cite{Romalis2002}, which operate at near zero fields, below $pT$, and at high alkali gas pressure that results in $\Gamma$ as small as $1$Hz. In addition to increasing the slope, $\frac{\partial \phi}{\partial B_{z}}$, development of alternate techniques to measure optical polarization rotation by improving $\delta \phi$, will lead to much higher sensitivities of the magnetometer.

    This article addresses the critical aspect of sensitive optical rotation angle measurement in response to magnetic field sensed by alkali atoms using the weak measurement method vis-\`a-vis the standard balanced optical polarimetry method. For demonstrating the technique, we use a magnetometer operating in the nonlinear regime of Faraday rotation  wherein the relaxation width is given by the width of the Bennett structure. Our attempt is to improve the sensitivity $\delta \phi$ via optical amplification, which is otherwise limited by the photon shot-noise of the detector. Standard balanced polarimetry method for optical rotation measurement is presented first, to compare it with results obtained using the weak measurement method. In the weak measurement method, horizontally polarized laser beam is passed through a Soleil-Babinet compensator (SBC) with fixed orientation and phase settings to generate optimized phase-polarization gradient in the beam cross-section. The state of polarization of the laser beam passing through the Rb atoms rotates in response to the weak applied magnetic field, which is measured as a shift in the beam centroid, after passing through an analyzer using a CMOS camera. 
\vspace*{-0.4cm}
\section{\label{sec:level3}Detection of Magneto-Optical Rotation}
The signal of the optical magnetometers is the rotation of the plane of polarization of the light beam after traversing the atomic medium subjected to longitudinal magnetic field. One of the simplest and widely used methods to measure this angle of rotation is balanced polarimetry. Though frequency\cite{Budker2006} and amplitude\cite{Gawlik2006} modulation techniques using pump-probe configuration give sensitivities larger than balanced polarimetry, we use non-modulating balanced polarimetry to compare the base sensitivity that can be achieved in a 'dc' measurement which can be further enhanced by using modulation techniques. 
\begin{figure}
\centering
\begin{tikzpicture}[scale=0.9]

\draw[thick, ->]   (-0.5,2)  -- (-0.5,1)  node[left] {$\hat{x}$};
\draw[thick, ->] (-0.5,2) --  (0.5,2) node[above] {$\hat{z}$};
\draw[] (-0.5,2) circle(0.12);
\draw[fill] (-0.5,2) circle(0.008);
\draw (-0.5,2) node[]{$\times$};
\draw (-0.5,2.1) node[above] {$\hat{y}$};

\draw[thick] (0,0) -- (3,0);
\draw[thick, ->] (3,0) -- (7.9,0);


 
\draw (-1,0.3) -- (-1,-0.3) -- node[below] {Laser}(0,-0.3) -- (0,0.3) -- (-1,0.3);

\draw (0.4,0.3) -- (0.4,-0.3) -- (0.6,-0.3) -- (0.6,0.3) -- node[above] {$\frac{\lambda}{2}$}(0.4,0.3);

\draw (1,0.3) -- (1,-0.3) -- (1.6,-0.3) -- (1.6,0.3) -- node[above] {PBS}(1,0.3);
\draw (1,0.3) -- (1.6,-0.3);

\draw[thick,->] (1.3,0) -- (1.3,-2) node[below] {SAS};

\draw (2,0.3) -- (2,-0.3) --  (2.2,-0.3) -- (2.2,0.3) -- node[above] {P} (2,0.3);

\draw (2.8,0.3) -- (2.8,-0.3) -- (3.2,-0.3) -- (3.2,0.3) -- node[above] {SBC}(2.8,0.3);

\draw (3.5,0.2) -- (3.9,0.2) -- (3.9,0.8) --  (6.1, 0.8) --  (6.1,0.2) -- (6.5,0.2);
\draw (3.5,0.3) -- (3.8,0.3) -- (3.8, 1.0) --  (6.2, 1.0) -- (6.2, 0.3) -- (6.5,0.3);
\draw (3.5,0.4) -- (3.7,0.4) -- (3.7, 1.2) --  (6.3, 1.2) -- (6.3, 0.4) -- (6.5,0.4);

\draw (3.5,-0.2) -- (3.9,-0.2) -- (3.9,-0.8) --    (6.1, -0.8) -- (6.1, -0.2) -- (6.5,-0.2);
\draw (3.5,-0.3) -- (3.8,-0.3)  -- (3.8, -1.0) --  (6.2, -1.0) -- (6.2, -0.3) -- (6.5,-0.3) ;
\draw (3.5,-0.4) -- (3.7,-0.4)   -- (3.7, -1.2)--  (6.3, -1.2) -- (6.3, -0.4) -- (6.5,-0.4);

\draw[->](4.1,2) -- node[above] {$\hat{B_{Z}}$} (5.9,2);

\draw[fill] (4.1,0.5)circle(0.05);
\draw[fill] (4.3,0.5)circle(0.05);
\draw[fill] (4.5,0.5)circle(0.05);
\draw[fill] (4.7,0.5)circle(0.05);
\draw[fill] (4.9,0.5)circle(0.05);
\draw[fill] (5.1,0.5)circle(0.05);
\draw[fill] (5.3,0.5)circle(0.05);
\draw[fill] (5.5,0.5)circle(0.05);
\draw[fill] (5.7,0.5)circle(0.05);
\draw[fill] (5.9,0.5)circle(0.05);

\draw[fill] (4.1,-0.5)circle(0.05);
\draw[fill] (4.3,-0.5)circle(0.05);
\draw[fill] (4.5,-0.5)circle(0.05);
\draw[fill] (4.7,-0.5)circle(0.05);
\draw[fill] (4.9,-0.5)circle(0.05);
\draw[fill] (5.1,-0.5)circle(0.05);
\draw[fill] (5.3,-0.5)circle(0.05);
\draw[fill] (5.5,-0.5)circle(0.05);
\draw[fill] (5.7,-0.5)circle(0.05);
\draw[fill] (5.9,-0.5)circle(0.05);

\draw (4.1,0.25) -- (4.1,-0.25) -- (5.9,-0.25) -- (5.9,0.25) -- (4.1,0.25);

\draw (6.9,0.3) -- (6.9,-0.3) -- (7.4,-0.3) -- (7.4,0.3) -- node[above] {A} (6.9,0.3);
\draw[] (6.95,0.3) -- (7.35,-0.3);

\draw[fill] (7.9,0.3) -- (7.9,-0.3) -- (8.1,-0.3) -- (8.1,0.3) -- node[above] {C/PD1} (7.9,0.3);

\draw[fill] (6.8,-0.8) --  (6.8,-1.0) -- (7.4,-1.0) node[below] {PD2} --  (7.4,-0.8) -- (6.8,-0.8);
\draw[->, thick] (7.15,0) -- (7.1,-0.8);

\draw[<-] (5.0,-0.28) -- (5.0, -1.5);
\draw (5.0,-1.5) -- (5.2,-1.5) node[right] {RB Cell};

\draw[<-] (4.9,-0.6) -- (4.9,-1.9);
\draw (4.9,-1.9) --  (5.2,-1.9) node[right] {Solenoid};

\draw[<-] (5.0, 1.2) -- (5.0, 1.5);
\draw (5.0,1.5) -- (5.2,1.5) node[right] {$\mu$ metal shield} ;
\end{tikzpicture}
\caption{\small{Schematic of experimental set-up used for balanced polarimetry and weak measurement methods. (PBS: Polarizing Beam-splitter, SAS: Saturation Absorption Spectroscopy, P: Polarizer (Glan-Taylor Polarizer), SBC: Soleil-Babinet Compensator (only for weak measurement), A: Analyzer (Glan-Taylor Polarizer), C: CMOS Camera, PD1 and PD2: Photo detectors (only for polarimetry measurement))}}
\label{Wm-Diagram}
\end{figure}
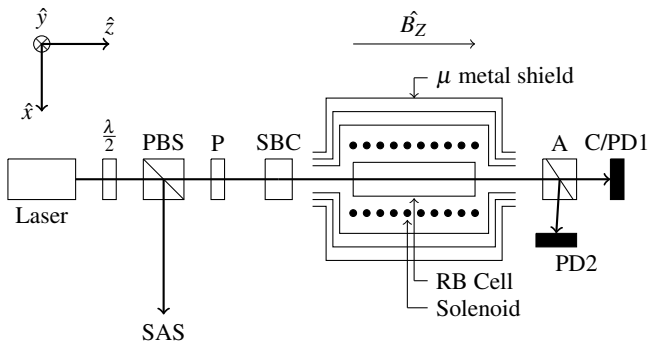

\vspace*{-5mm}
\subsection{\label{sec:level4}Balanced Polarimetry Method}
In the balanced polarimetry method, intensity of the horizontal (H) and vertical (V) linear polarization components of the light beam propagating through the medium are measured simultaneously using two photo-detectors and the magneto-optical rotation, for small rotations is calculated using the formula, 
$\phi = \frac{1}{2} \frac{I(H) - I(V)}{I(H) + I(V)} $, 
where $I(H)$ and $I(V)$ are the intensity of horizontal $\mathbf{\hat{x}}$ and vertical $\mathbf{\hat{y}}$ polarization components respectively. The theoretically achievable detection sensitivity in this scheme is limited by the photon shot-noise \cite{Grangier1987}, which is calculated to be 46 $\mu rad$ for our experimental system. In our measurement (experiment schematic shown in FIG.~\ref{Wm-Diagram} without the SBC), the state-of-polarization of the on-resonance light beam ($\lambda$ = 780nm) from the diode laser (Vantage 7100, NewFocus) entering the atomic medium (Rb) is kept fixed at 0.785 rad ($45^{o}$) with respect to $\mathbf{\hat{x}}$ using the polarizer P. The Rubidium cell is 7 cm long and 2.5 cm diameter and is placed on-axis within the solenoid coils surrounded by a 3-layer $\mu$-metal shield and known longitudinal magnetic field is applied by passing current from a stabilized current source (Keithley 6221). The frequency of the laser is stabilized using saturation absorption spectroscopy \cite{John2008} to be at the D2 transition of Rb$^{85}$ atoms. The laser beam is nearly coolimated with a power of 1mW over a beam diameter of 0.4cm, corresponding to beam intensity of 8mW/cm$^{2}$ at the vapour cell. Light exiting the Rb cell  passes through Glan-Taylor analyzer A whose transmission axis is parallel to $\mathbf{\hat{x}}$, such that the horizontal component of polarization passes though it and falls on the photo-detector, PD1, while the vertical component is reflected and falls on the photo-detector PD2. In the absence of applied magnetic field, equal intensities are measured in the two photo-detectors (PD1, PD2), corresponding to zero magneto-optical rotation of the plane of polarization of light exiting the atomic medium. When the magnetic field is turned on, the plane of polarization of the linearly polarized light leaving the atomic medium rotates, resulting in slightly different intensities measured in the two detectors, giving rise to magneto-optical rotation signal calculated. Our experimental measurement is shown in Fig. 2, along with the fit using Eqn.~\ref{mo-rotation} giving $\Gamma = 1$~MHz and $l/l_o =$ 0.2. The slope of the linear region near zero magnetic field is calculated to be 18.3 $mrad/\mu T$. From the slope, the sensitivity $\delta B_{z}$, of the magentometer is calculated to be 2.5 nT. However, upon close observation, it can be seen from FIG. 2 that even around zero magnetic field, one sees departure of the experimental data from the theoretical curve, indicating potential limitation of the balanced polarimetry method in measuring small rotation angles, corresponding to small value of applied magnetic field.

\vspace*{-5mm}
\subsection{\label{sec:level5}Weak Measurement Method}
One of the objectives of this article is to demonstrate optical magnetometry using weak measurement method and its capability to measure polarization rotation angles and hence the applied magnetic field smaller than the measurable limit of the balanced polarimetry method. The high sensitivity of the weak measurement method used for obtaining the polarization rotation angle, is achieved via optical amplification based on the concept introduced by Aharonov et al. \cite{Aharonov1988} using the Stern-Gerlach device.
\begin{figure}
\centering
\includegraphics[width=7cm]{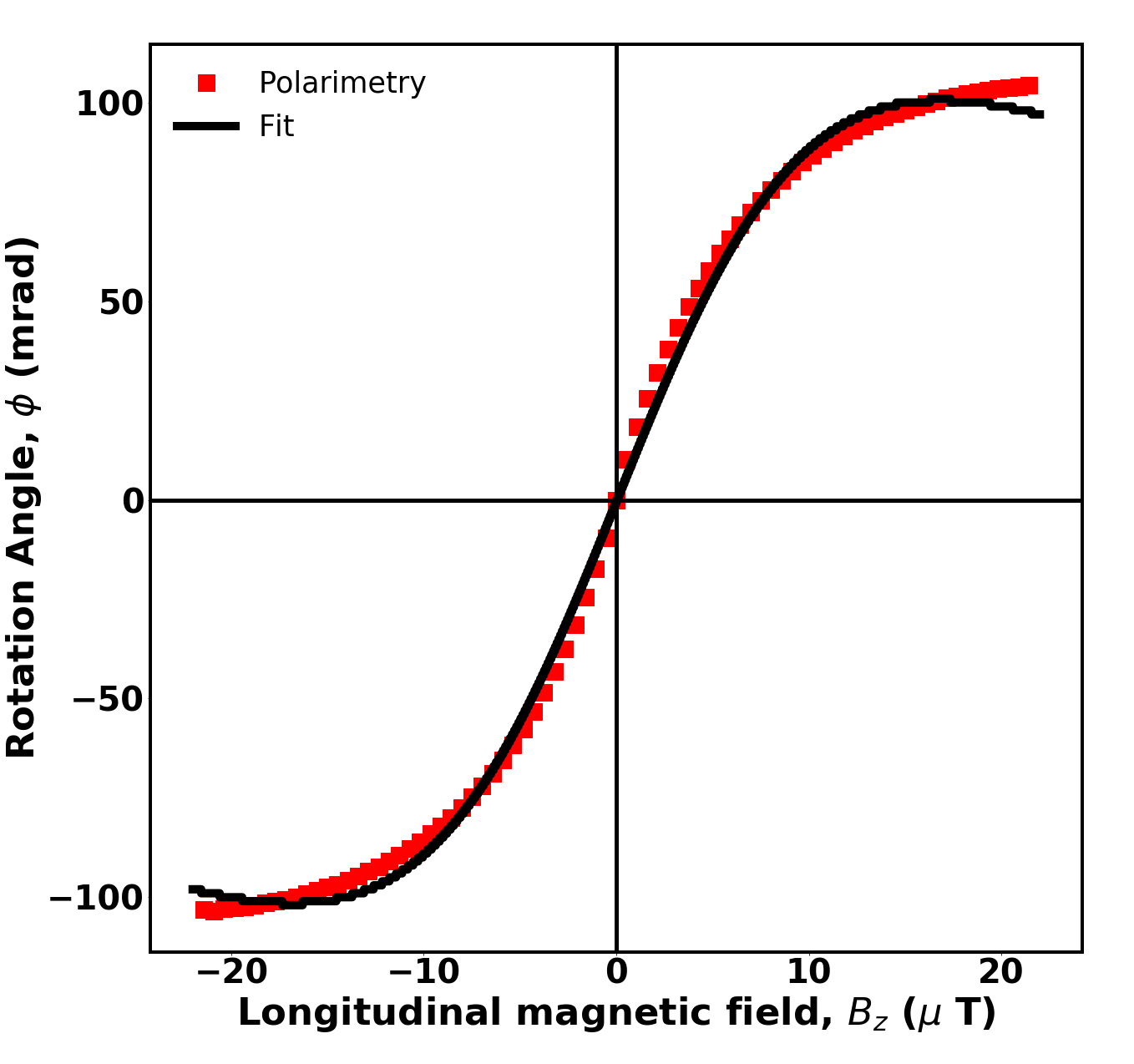}
\caption{\small{Rotation angle measured as a function of longitudinal magnetic field using balanced polarimetry method and theoretical fit}}
\label{bp-plot}
\end{figure}
 We use this concept implemented in classical systems \cite{Sudarshan1989,Ritchie1991} for amplifying weak effects to measure small rotation in the state of polarization in response to applied magnetic field, leading to a large change in the Gaussian beam intensity profile. This is realized by coupling the state of polarization and the transverse momentum degree of freedom of the input Gaussian beam using a Soleil-Babinet compensator (SBC). Here we refer the state of polarization of the Gaussian beam as the observable and its transverse momentum vector with canonical conjugate transverse position as the apparatus. The SBC is used to couple the polarization state and the momentum vector resulting in inhomogeneously polarized Gaussian beam, also known as spin-orbit beam and has been used to measure weak optical chirality \cite{rashmi1,rashmi2}.

\begin{figure}
\centering
\includegraphics[width=9cm, height=7cm]{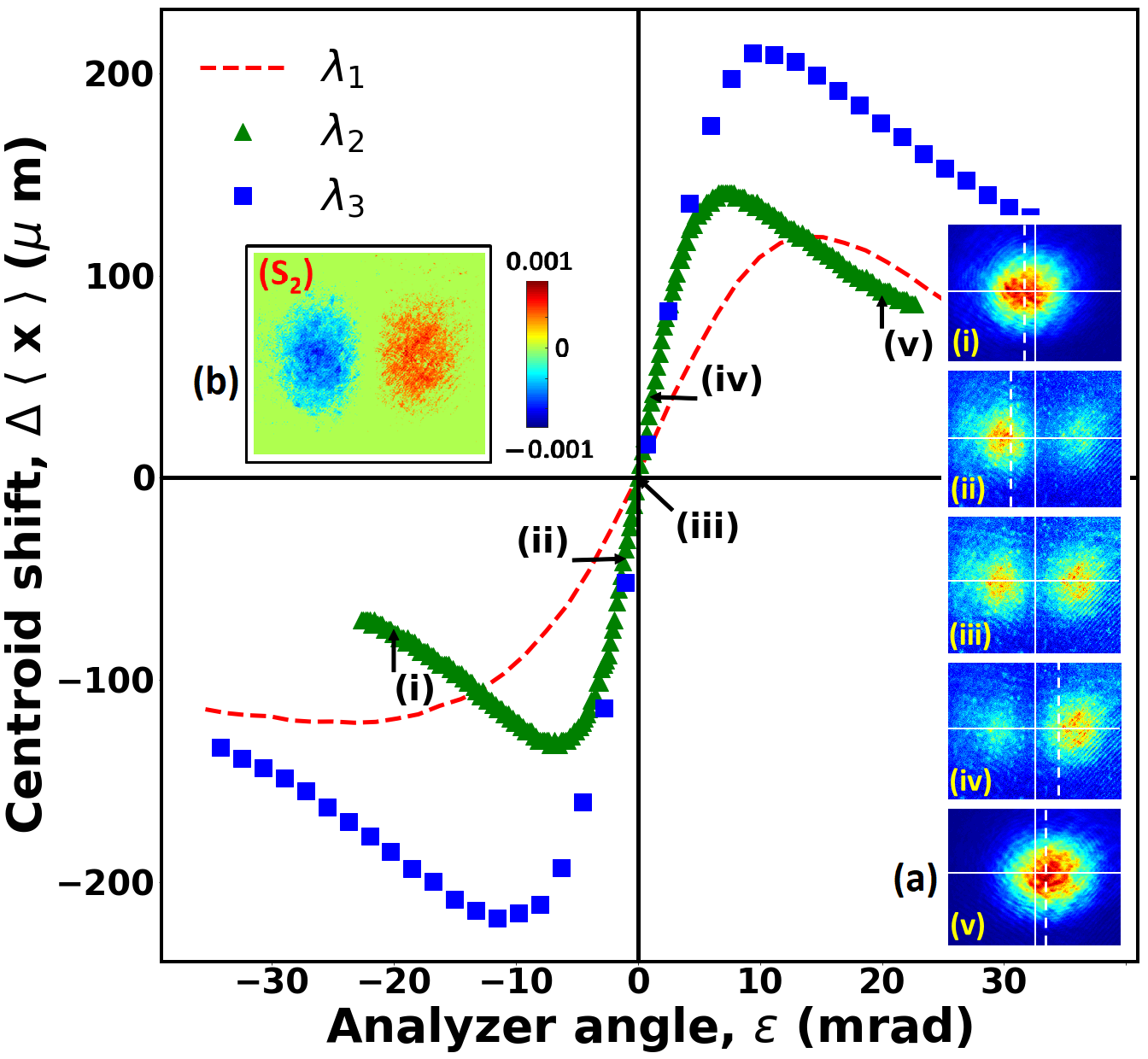}
\caption{\small{Weak measurement calibration plot of beam centroid shift versus analyzer angle ($\lambda_{1}$, $\lambda_{2}$,  $\lambda_{3}$ corresponds to ($\theta$, $\alpha$, $\beta_{o}$) = ($0mrad$, $174.5mrad$, $43 mrad$), ($61.1mrad$, $0mrad$, $38 mrad$), ($0mrad$, $17.4mrad$, $0.1 mrad$) and corresponding slope of the linear portion are $11.6 \mu m/mrad$, $35 \mu m/mrad$, $38 \mu m/mrad$ respectively). Inset shows (a) the beam profiles corresponding to the labeled positions on the plot and (b) experimentally measured Stokes parameter $S_{2}$ .}}
\label{Wm-SBc}
\end{figure}

In our experimental setup shown in FIG.~\ref{Wm-Diagram}, the polarizer P pre-selects the state of polarization of the laser beam with Gaussian intensity profile to be at an angle $\alpha$ with respect to $\mathbf{\hat{x}}$. The transverse electric field of the pre-selected Gaussian beam is given by, 
\begin{align}
E_{i} = 
\begin{bmatrix}
E_{o} \cos(\alpha) \\
E_{o} \sin(\alpha)
\end{bmatrix}
\exp\left[ -(x^{2} + y^{2})/\omega_{o}^{2} \right]
\label{pre-electricfield}
\end{align}
 where $\omega_{o}$ is the beam waist and $E_o$ is the electric field amplitude. Passing through the Soleil-Babinet compensator (SBC) \cite{Max1997}, a zero-order variable phase retarder, whose fast-axis is oriented at an angle $\theta$ with respect to $\mathbf{\hat{x}}$ results in the appearance of phase-polarization gradient across the paraxial Gaussian beam. The Jones matrix of the SBC can be written as,
\begin{align}
J_{SBC} = \begin{bmatrix}
\exp\left[i \frac{\Phi_{o}}{2} - \frac{\chi}{2 z_{R}} x \right] & 0 \\
0 & \exp\left[-i \frac{\Phi_{o}}{2} + \frac{\chi}{2 z_{R}} x \right]
\end{bmatrix}
\label{SBC-Jones}
\end{align}
where $\Phi_{o}$ is the phase difference between the ordinary and extra-ordinary rays corresponding to the beam center, $\chi$ is the phase gradient across the beam cross-section and $z_{R}$ is the Rayleigh range. The analyzer A, post-selects the orthogonal polarization state, $\beta = \alpha + \frac{\pi}{2}$, resulting in transverse electric field,
\begin{align}
E_{f} = 
\begin{bmatrix}
\cos(\beta)^{2} & \cos(\beta) \sin(\beta) \\
\cos(\beta) \sin(\beta) & \sin(\beta)^{2}
\end{bmatrix}
R(\theta)^{-1}J_{SBC}R(\theta)E_{i}
\label{post-electricfield}
\end{align}
where $R(\theta)$ is the rotation matrix. A CMOS-camera measures the beam intensity profile after passing though the analyzer. When the analyzer is oriented nearly orthogonal ($\beta - \beta_{o}$) to the polarizer we measure minimum output intensity and a two-lobe pattern, resembling Hermite-Gaussian ($\mathrm{HG}_{10}$) mode at the detector. $\beta_{0}$ is the offset angle of the analyzer dialed to remove other phase contributions. The centroid of the intensity profile is calculated using
\begin{align}
\braket{x} = \frac{ \int x I(x,y) dx dy}{\int I(x,y)dx dy} \quad \braket{y}= \frac{ \int y I(x,y) dx dy}{\int I(x,y)dx dy}
\label{centroid}
\end{align}
where $I(x,y)$ is the beam intensity given by, $I(x,y) = E_{f_{x}}E_{f_{x}}^{*}+ E_{f_{y}}E_{f_{y}}^{*}$. Changing the analyzer angle by a small amount $\varepsilon$ from $(\beta - \beta_{o})$, will distort the beam profile at the detector from the $\mathrm{HG}_{10}$ mode \ref{Wm-SBc}. The resulting centroid shift, defined as $\Delta\braket{x} = \braket{x_{\varepsilon}}-\braket{x_{\varepsilon=0}}$ (where $x_{\epsilon}$ is the x-coordinate of the beam centroid corresponding to angle $\varepsilon$) is measured at various analyzer angle $\varepsilon$ and for different SBC orientation $\theta$ and incident polarization $\alpha$ as shown in FIG.~\ref{Wm-SBc}. A calibration plot is made between the centroid shift $\Delta\braket{x}$ and $\varepsilon$, which shows a linear behavior around the post-selected angle of the analyzer  $(\beta-\beta_{o})$, shown in FIG.~\ref{Wm-SBc}. The centroid shift $\Delta\braket{y}$ was found to be zero for orthogonal polarizer-analyzer orientations.  The optical beam generated by propagating through the SBC possesses phase gradient along the x-axis and is confirmed by measuring the Stokes parameter,$S_2$\cite{Dennis2010}. Inset in FIG.~\ref{Wm-SBc} shows the weak orientation value of the state of polarization of the two lobes of the HG mode indicating clearly the presence of phase-polarization gradient in the beam cross-section.

\begin{figure}
\centering
\includegraphics[width=7cm]{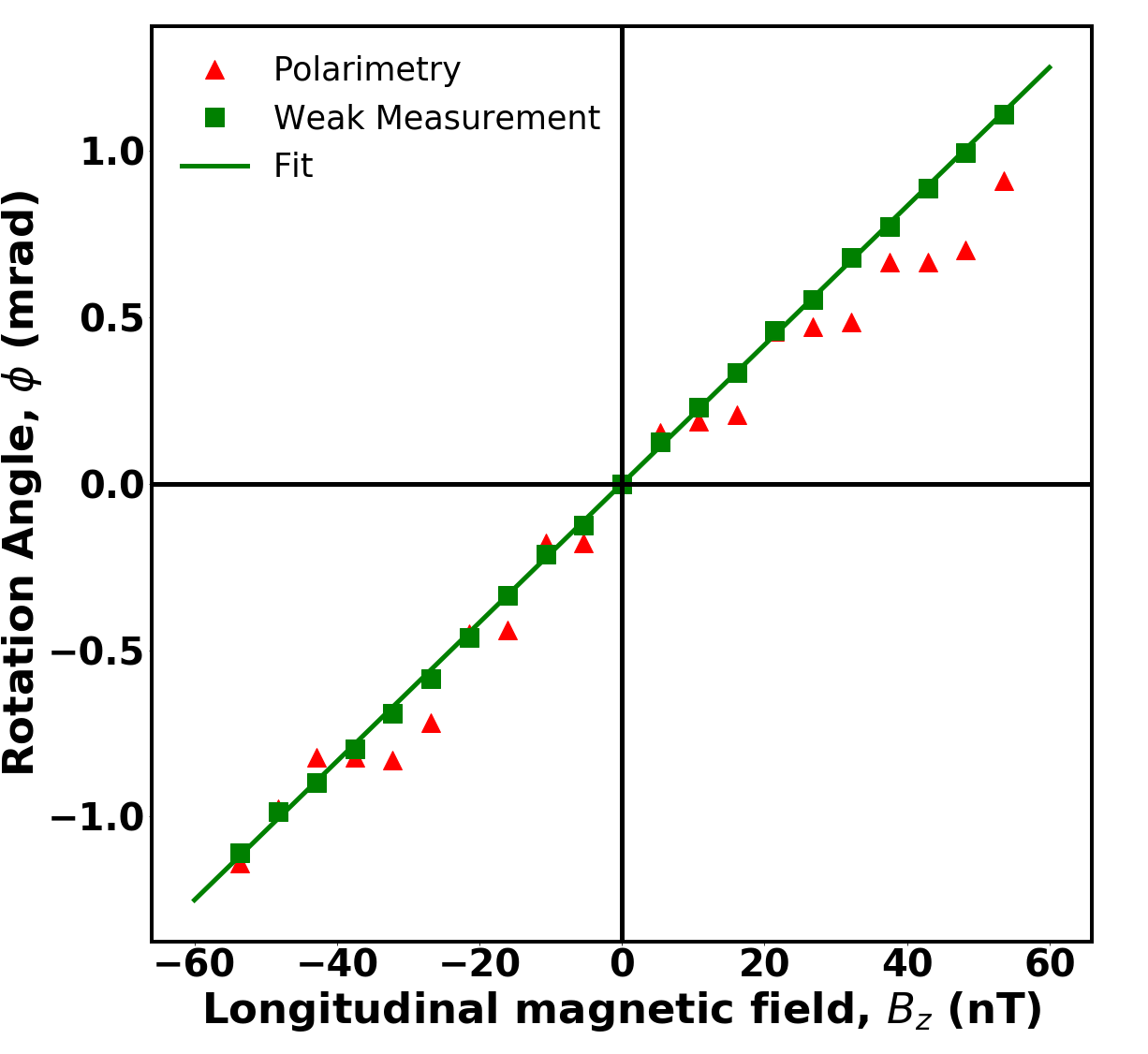}
\caption{\small{Rotation angle measured as a function of longitudinal magnetic field using balanced polarimetry and WM methods}}.
\label{Wm-Rotation}
\end{figure}

 The linear portion of the beam centroid shift $\Delta\braket{x}$ versus $\varepsilon$ calibration plot is used to calculate the magneto-optical rotation of the plane of polarization of light exiting the atomic medium. When the analyzer is brought back to $\beta - \beta_{o}$, in the absence of the magnetic field, the medium does not affect the state of polarization and we get a $\mathrm{HG}_{10}$ like mode at the CMOS detector. When the magnetic field is turned on, the plane of polarization of the Gaussian beam changes due to resonant light-atom interaction, leading to a change in the intensity structure of the output beam due to change in the post-selection state due to magneto-optical rotation. From the measured intensity profile, the centroid shift is calculated for various values of the applied magnetic field $B_{z}$. The experiment was carried out with the SBC oriented at $\theta=61.1 ~mrad$ and for horizontally polarized input beam ($\alpha = 0 ~mrad$). Without the Rb$^{85}$ gas cell in the beam path, the analyzer performs post-selection process resulting in the calibration graph shown in Fig.~\ref{Wm-SBc} (green triangle plot, with the slope of linear region = $35\mu m/mrad$). However, the introduction  of Rb$^{85}$ cell in the magnetic field, with the analyzer kept fixed at the crossed position results in a change in the behavior of the output mode pattern due to magneto-optical rotation by angle $\phi$.  The resulting beam centroid shift is used to calculate the optical rotation angle. It is to be noted that $\Delta\braket{y}=0$ for the present case. Using the linear region of the calibration graph shown in FIG.~\ref{Wm-SBc}, corresponding to $\theta = 61.1~mrad$ and $\alpha = 0~mrad$, the magneto-optical rotation of the plane of polarization of light exiting the atomic medium with respect to the applied longitudinal magnetic field is calculated. The rotation angle $\phi$ with respect to longitudinal magnetic field $B_{z}$ measured using weak measurement method are shown in FIG.~\ref{Wm-Rotation} along with the data from balanced polarimetry. The data from weak measurement is fitted with Eqn. \ref{mo-rotation} and the slope is found to be 20.8$mrad/\mu T$. One can easily see from the figure that the magneto-optical rotation measured using the weak measurement method does not show jumps due to resolution limitation like the balanced polarimetry method.
 
 As can be seen from FIG.~\ref{Wm-SBc}, the slope of the linear region of the graph and hence the sensitivity of the method depends on the ($\theta$ and $\alpha$) parameters of the SBC and can be tuned over a range of values. However, the highest achievable sensitivity to magneto-optical rotation by weak measurement is determined by the  shot-noise limited centroid shift that can be measured, which is 0.56$\mu$m for our detector, leading to a sensitivity in optical rotation,  $\delta\phi$ of  16 $\mu rad$. From the slope for the weak measurement shown in FIG.~\ref{Wm-Rotation} and {$\delta\phi$}, the sensitivity $\delta B_{z}$ to the longitudinal magnetic field is calculated to be 0.8 nT. 
 It is important to note that the range of rotation angle that we can measure using this method is limited to the linear region of the centroid shift. However, for measuring magnetic field corresponding to rotation beyond the linear region of the centroid shift, possible options will be to offset zero crossing and hence the measurement range by either applying a known dc field and / or rotating the analyzer angle to bring the detection to within the linear measurement range. 

\vspace*{-5mm}
\section{\label{sec:level7}Conclusion}
We proposed and demonstrated weak measurement method for ultra-sensitive, single-beam atom-optical magnetometer whose sensitivity $\delta\phi$ to magneto-optical rotation is given by,
$
\delta\phi = \left(\frac{\partial}{\partial \phi}\Delta\braket{x} \right)^{-1} \delta\Delta\braket{x}
$,
where $\frac{\partial}{\partial \phi}\Delta\braket{x}$ is the slope of centroid shift with respect to analyser angle $\varepsilon$ and $\delta \Delta \braket{x}$ is the sensitivity to centroid shift measured by CMOS camera which is limited by photon shot noise of the camera. Unlike measurement of $\delta\phi$ using balanced polarimetry method, which is totally photon shot noise dependent, in the weak measurement method we have an additional control, the slope of the centroid shift $\frac{\partial}{\partial \phi}\Delta\braket{x}$, to improve the sensitivity $\delta\phi$ to magneto-optical rotation. It was observed that the slope increases with decreasing the relative angle between the fast axis of SBC and the initial state of polarization of the light beam, and the best slope measured is $38\mu m/mrad$ and the corresponding sensitivity is $\delta\phi$ is 16 $\mu rad$. Currently the magnetometer works in the non-linear Faraday regime, with a dynamic range of 0.5 $\mu$T, and has sensitivity to longitudinal magnetic field of $\delta B_{z}$ $\approx$ 1 nT. Employing SERF regime is expected to reduce $\Gamma$ to a few Hz, enhancing the magnetic field sensitivity\cite{Budker2008,Romalis2009}, $\frac{\partial \phi}{\partial B_{z}}$ by $10^6$ while reducing its dynamic range. Coupling this with increased sensitivity of the weak measurement technique, better sensitivity $\delta B_{z}$ of the order of $\sim$ f T can be achieved. The weak measurement method is a compact single beam dc technique that does not involve any modulation in optical magnetometers, making it  very useful over a broad band spectral range for optical magnetic sensors especially the compact, MEMS\cite{Sheng2016,Sheng2017} sensors, complimenting other modulation and radio frequency based techniques.

\vspace*{-5mm}
\section*{\label{sec:level8}Acknowledgement}
\vspace*{-4mm}
RRS thanks UGC-RGNF for research fellowship, NKV thanks  Science and Engineering Research Board (SERB), Department of Science and Technology (DST), India for continued financial support to this area of research, GR thanks Defense Research and Development Organisation, India for financial assistance. One of the authors, RRS passed away under unfortunate circumstances in the final stages of this work and this article is dedicated to his memory.
\section*{\label{sec:level9}References}

\bibliography{References}

\begin{thebibliography}{28}%
\makeatletter
\providecommand \@ifxundefined [1]{%
 \@ifx{#1\undefined}
}%
\providecommand \@ifnum [1]{%
 \ifnum #1\expandafter \@firstoftwo
 \else \expandafter \@secondoftwo
 \fi
}%
\providecommand \@ifx [1]{%
 \ifx #1\expandafter \@firstoftwo
 \else \expandafter \@secondoftwo
 \fi
}%
\providecommand \natexlab [1]{#1}%
\providecommand \enquote  [1]{``#1''}%
\providecommand \bibnamefont  [1]{#1}%
\providecommand \bibfnamefont [1]{#1}%
\providecommand \citenamefont [1]{#1}%
\providecommand \href@noop [0]{\@secondoftwo}%
\providecommand \href [0]{\begingroup \@sanitize@url \@href}%
\providecommand \@href[1]{\@@startlink{#1}\@@href}%
\providecommand \@@href[1]{\endgroup#1\@@endlink}%
\providecommand \@sanitize@url [0]{\catcode `\\12\catcode `\$12\catcode
  `\&12\catcode `\#12\catcode `\^12\catcode `\_12\catcode `\%12\relax}%
\providecommand \@@startlink[1]{}%
\providecommand \@@endlink[0]{}%
\providecommand \url  [0]{\begingroup\@sanitize@url \@url }%
\providecommand \@url [1]{\endgroup\@href {#1}{\urlprefix }}%
\providecommand \urlprefix  [0]{URL }%
\providecommand \Eprint [0]{\href }%
\providecommand \doibase [0]{http://dx.doi.org/}%
\providecommand \selectlanguage [0]{\@gobble}%
\providecommand \bibinfo  [0]{\@secondoftwo}%
\providecommand \bibfield  [0]{\@secondoftwo}%
\providecommand \translation [1]{[#1]}%
\providecommand \BibitemOpen [0]{}%
\providecommand \bibitemStop [0]{}%
\providecommand \bibitemNoStop [0]{.\EOS\space}%
\providecommand \EOS [0]{\spacefactor3000\relax}%
\providecommand \BibitemShut  [1]{\csname bibitem#1\endcsname}%
\let\auto@bib@innerbib\@empty
\bibitem [{\citenamefont {Budker}\ \emph
  {et~al.}(2002{\natexlab{a}})\citenamefont {Budker}, \citenamefont {Gawlik},
  \citenamefont {Kimball}, \citenamefont {Rochester}, \citenamefont
  {Yashchuk},\ and\ \citenamefont {Weis}}]{Budker2002}%
  \BibitemOpen
  \bibfield  {author} {\bibinfo {author} {\bibfnamefont {D.}~\bibnamefont
  {Budker}}, \bibinfo {author} {\bibfnamefont {W.}~\bibnamefont {Gawlik}},
  \bibinfo {author} {\bibfnamefont {D.~F.}\ \bibnamefont {Kimball}}, \bibinfo
  {author} {\bibfnamefont {S.~M.}\ \bibnamefont {Rochester}}, \bibinfo {author}
  {\bibfnamefont {V.~V.}\ \bibnamefont {Yashchuk}}, \ and\ \bibinfo {author}
  {\bibfnamefont {A.}~\bibnamefont {Weis}},\ }\href {\doibase
  10.1103/RevModPhys.74.1153} {\bibfield  {journal} {\bibinfo  {journal} {Rev.
  Mod. Phys.}\ }\textbf {\bibinfo {volume} {74}},\ \bibinfo {pages} {1153}
  (\bibinfo {year} {2002}{\natexlab{a}})}\BibitemShut {NoStop}%
\bibitem [{\citenamefont {Budker}\ and\ \citenamefont
  {Romalis}(2007)}]{Budker2007}%
  \BibitemOpen
  \bibfield  {author} {\bibinfo {author} {\bibfnamefont {D.}~\bibnamefont
  {Budker}}\ and\ \bibinfo {author} {\bibfnamefont {M.}~\bibnamefont
  {Romalis}},\ }\href {\doibase 10.1038/nphys566} {\bibfield  {journal}
  {\bibinfo  {journal} {Nature Physics}\ }\textbf {\bibinfo {volume} {3}}
  (\bibinfo {year} {2007}),\ 10.1038/nphys566}\BibitemShut {NoStop}%
\bibitem [{\citenamefont {Taylor}\ \emph {et~al.}(2008)\citenamefont {Taylor},
  \citenamefont {Cappellaro}, \citenamefont {Childress}, \citenamefont {Jiang},
  \citenamefont {Budker}, \citenamefont {Hemmer}, \citenamefont {Yacoby},
  \citenamefont {Walsworth},\ and\ \citenamefont {Lukin}}]{Taylor2008}%
  \BibitemOpen
  \bibfield  {author} {\bibinfo {author} {\bibfnamefont {J.~M.}\ \bibnamefont
  {Taylor}}, \bibinfo {author} {\bibfnamefont {P.}~\bibnamefont {Cappellaro}},
  \bibinfo {author} {\bibfnamefont {L.}~\bibnamefont {Childress}}, \bibinfo
  {author} {\bibfnamefont {L.}~\bibnamefont {Jiang}}, \bibinfo {author}
  {\bibfnamefont {D.}~\bibnamefont {Budker}}, \bibinfo {author} {\bibfnamefont
  {P.~R.}\ \bibnamefont {Hemmer}}, \bibinfo {author} {\bibfnamefont
  {A.}~\bibnamefont {Yacoby}}, \bibinfo {author} {\bibfnamefont
  {R.}~\bibnamefont {Walsworth}}, \ and\ \bibinfo {author} {\bibfnamefont
  {M.~D.}\ \bibnamefont {Lukin}},\ }\href {\doibase 10.1038/nphys1075}
  {\bibfield  {journal} {\bibinfo  {journal} {Nature Physics}\ }\textbf
  {\bibinfo {volume} {4}} (\bibinfo {year} {2008}),\
  10.1038/nphys1075}\BibitemShut {NoStop}%
\bibitem [{\citenamefont {Dang}\ \emph {et~al.}(2010)\citenamefont {Dang},
  \citenamefont {Maloof},\ and\ \citenamefont {Romalis}}]{Dang2010}%
  \BibitemOpen
  \bibfield  {author} {\bibinfo {author} {\bibfnamefont {H.~B.}\ \bibnamefont
  {Dang}}, \bibinfo {author} {\bibfnamefont {A.~C.}\ \bibnamefont {Maloof}}, \
  and\ \bibinfo {author} {\bibfnamefont {M.~V.}\ \bibnamefont {Romalis}},\
  }\href {\doibase 10.1063/1.3491215} {\bibfield  {journal} {\bibinfo
  {journal} {Applied Physics Letters}\ }\textbf {\bibinfo {volume} {97}},\
  \bibinfo {pages} {151110} (\bibinfo {year} {2010})},\ \Eprint
  {http://arxiv.org/abs/https://doi.org/10.1063/1.3491215}
  {https://doi.org/10.1063/1.3491215} \BibitemShut {NoStop}%
\bibitem [{\citenamefont {Fan}\ \emph {et~al.}(2018)\citenamefont {Fan},
  \citenamefont {Zhang}, \citenamefont {Yang}, \citenamefont {Cui},
  \citenamefont {Zhou},\ and\ \citenamefont {Feng}}]{Fan2018}%
  \BibitemOpen
  \bibfield  {author} {\bibinfo {author} {\bibfnamefont {T.}~\bibnamefont
  {Fan}}, \bibinfo {author} {\bibfnamefont {L.}~\bibnamefont {Zhang}}, \bibinfo
  {author} {\bibfnamefont {X.}~\bibnamefont {Yang}}, \bibinfo {author}
  {\bibfnamefont {S.}~\bibnamefont {Cui}}, \bibinfo {author} {\bibfnamefont
  {T.}~\bibnamefont {Zhou}}, \ and\ \bibinfo {author} {\bibfnamefont
  {Y.}~\bibnamefont {Feng}},\ }\href {\doibase 10.1364/OL.43.000001} {\bibfield
   {journal} {\bibinfo  {journal} {Opt. Lett.}\ }\textbf {\bibinfo {volume}
  {43}},\ \bibinfo {pages} {1} (\bibinfo {year} {2018})}\BibitemShut {NoStop}%
\bibitem [{\citenamefont {Macaluso}\ and\ \citenamefont
  {Corbino}(1898)}]{Macaluso1898}%
  \BibitemOpen
  \bibfield  {author} {\bibinfo {author} {\bibfnamefont {D.}~\bibnamefont
  {Macaluso}}\ and\ \bibinfo {author} {\bibfnamefont {O.~M.}\ \bibnamefont
  {Corbino}},\ }\href@noop {} {\bibfield  {journal} {\bibinfo  {journal} {Nuovo
  Cimento}\ }\textbf {\bibinfo {volume} {8}},\ \bibinfo {pages} {257} (\bibinfo
  {year} {1898})}\BibitemShut {NoStop}%
\bibitem [{\citenamefont {Macaluso}\ and\ \citenamefont
  {Corbino}(1899)}]{Macaluso1899}%
  \BibitemOpen
  \bibfield  {author} {\bibinfo {author} {\bibfnamefont {D.}~\bibnamefont
  {Macaluso}}\ and\ \bibinfo {author} {\bibfnamefont {O.~M.}\ \bibnamefont
  {Corbino}},\ }\href@noop {} {\bibfield  {journal} {\bibinfo  {journal} {Nuovo
  Cimento}\ }\textbf {\bibinfo {volume} {9}},\ \bibinfo {pages} {384} (\bibinfo
  {year} {1899})}\BibitemShut {NoStop}%
\bibitem [{\citenamefont {Budker}\ and\ \citenamefont
  {Kimball}(2013)}]{Budker2013}%
  \BibitemOpen
  \bibinfo {editor} {\bibfnamefont {D.}~\bibnamefont {Budker}}\ and\ \bibinfo
  {editor} {\bibfnamefont {D.~F.~J.}\ \bibnamefont {Kimball}},\ eds.,\
  \href@noop {} {\emph {\bibinfo {title} {Optical Magnetometry}}},\ \bibinfo
  {edition} {1st}\ ed.\ (\bibinfo  {publisher} {Cambridge University Press},\
  \bibinfo {year} {2013})\BibitemShut {NoStop}%
\bibitem [{\citenamefont {Bennett}(1962)}]{Bennett1962}%
  \BibitemOpen
  \bibfield  {author} {\bibinfo {author} {\bibfnamefont {W.~R.}\ \bibnamefont
  {Bennett}},\ }\href {\doibase 10.1103/PhysRev.126.580} {\bibfield  {journal}
  {\bibinfo  {journal} {Phys. Rev.}\ }\textbf {\bibinfo {volume} {126}},\
  \bibinfo {pages} {580} (\bibinfo {year} {1962})}\BibitemShut {NoStop}%
\bibitem [{\citenamefont {Budker}\ \emph
  {et~al.}(2002{\natexlab{b}})\citenamefont {Budker}, \citenamefont {Kimball},
  \citenamefont {Rochester},\ and\ \citenamefont {Yashchuk}}]{Kimball2002}%
  \BibitemOpen
  \bibfield  {author} {\bibinfo {author} {\bibfnamefont {D.}~\bibnamefont
  {Budker}}, \bibinfo {author} {\bibfnamefont {D.~F.}\ \bibnamefont {Kimball}},
  \bibinfo {author} {\bibfnamefont {S.~M.}\ \bibnamefont {Rochester}}, \ and\
  \bibinfo {author} {\bibfnamefont {V.~V.}\ \bibnamefont {Yashchuk}},\ }\href
  {\doibase 10.1103/PhysRevA.65.033401} {\bibfield  {journal} {\bibinfo
  {journal} {Phys. Rev. A}\ }\textbf {\bibinfo {volume} {65}},\ \bibinfo
  {pages} {033401} (\bibinfo {year} {2002}{\natexlab{b}})}\BibitemShut
  {NoStop}%
\bibitem [{\citenamefont {Budker}\ \emph {et~al.}(1998)\citenamefont {Budker},
  \citenamefont {Yashchuk},\ and\ \citenamefont {Zolotorev}}]{Budker1998}%
  \BibitemOpen
  \bibfield  {author} {\bibinfo {author} {\bibfnamefont {D.}~\bibnamefont
  {Budker}}, \bibinfo {author} {\bibfnamefont {V.}~\bibnamefont {Yashchuk}}, \
  and\ \bibinfo {author} {\bibfnamefont {M.}~\bibnamefont {Zolotorev}},\ }\href
  {\doibase 10.1103/PhysRevLett.81.5788} {\bibfield  {journal} {\bibinfo
  {journal} {Phys. Rev. Lett.}\ }\textbf {\bibinfo {volume} {81}},\ \bibinfo
  {pages} {5788} (\bibinfo {year} {1998})}\BibitemShut {NoStop}%
\bibitem [{\citenamefont {Budker}\ \emph {et~al.}(2000)\citenamefont {Budker},
  \citenamefont {Kimball}, \citenamefont {Rochester}, \citenamefont
  {Yashchuk},\ and\ \citenamefont {Zolotorev}}]{Budker2000}%
  \BibitemOpen
  \bibfield  {author} {\bibinfo {author} {\bibfnamefont {D.}~\bibnamefont
  {Budker}}, \bibinfo {author} {\bibfnamefont {D.~F.}\ \bibnamefont {Kimball}},
  \bibinfo {author} {\bibfnamefont {S.~M.}\ \bibnamefont {Rochester}}, \bibinfo
  {author} {\bibfnamefont {V.~V.}\ \bibnamefont {Yashchuk}}, \ and\ \bibinfo
  {author} {\bibfnamefont {M.}~\bibnamefont {Zolotorev}},\ }\href {\doibase
  10.1103/PhysRevA.62.043403} {\bibfield  {journal} {\bibinfo  {journal} {Phys.
  Rev. A}\ }\textbf {\bibinfo {volume} {62}},\ \bibinfo {pages} {043403}
  (\bibinfo {year} {2000})}\BibitemShut {NoStop}%
\bibitem [{\citenamefont {Allred}\ \emph {et~al.}(2002)\citenamefont {Allred},
  \citenamefont {Lyman}, \citenamefont {Kornack},\ and\ \citenamefont
  {Romalis}}]{Romalis2002}%
  \BibitemOpen
  \bibfield  {author} {\bibinfo {author} {\bibfnamefont {J.~C.}\ \bibnamefont
  {Allred}}, \bibinfo {author} {\bibfnamefont {R.~N.}\ \bibnamefont {Lyman}},
  \bibinfo {author} {\bibfnamefont {T.~W.}\ \bibnamefont {Kornack}}, \ and\
  \bibinfo {author} {\bibfnamefont {M.~V.}\ \bibnamefont {Romalis}},\ }\href
  {\doibase 10.1103/PhysRevLett.89.130801} {\bibfield  {journal} {\bibinfo
  {journal} {Phys. Rev. Lett.}\ }\textbf {\bibinfo {volume} {89}},\ \bibinfo
  {pages} {130801} (\bibinfo {year} {2002})}\BibitemShut {NoStop}%
\bibitem [{\citenamefont {Pustelny}\ \emph {et~al.}(2006)\citenamefont
  {Pustelny}, \citenamefont {Jackson~Kimball}, \citenamefont {Rochester},
  \citenamefont {Yashchuk}, \citenamefont {Gawlik},\ and\ \citenamefont
  {Budker}}]{Budker2006}%
  \BibitemOpen
  \bibfield  {author} {\bibinfo {author} {\bibfnamefont {S.}~\bibnamefont
  {Pustelny}}, \bibinfo {author} {\bibfnamefont {D.~F.}\ \bibnamefont
  {Jackson~Kimball}}, \bibinfo {author} {\bibfnamefont {S.~M.}\ \bibnamefont
  {Rochester}}, \bibinfo {author} {\bibfnamefont {V.~V.}\ \bibnamefont
  {Yashchuk}}, \bibinfo {author} {\bibfnamefont {W.}~\bibnamefont {Gawlik}}, \
  and\ \bibinfo {author} {\bibfnamefont {D.}~\bibnamefont {Budker}},\ }\href
  {\doibase 10.1103/PhysRevA.73.023817} {\bibfield  {journal} {\bibinfo
  {journal} {Phys. Rev. A}\ }\textbf {\bibinfo {volume} {73}},\ \bibinfo
  {pages} {023817} (\bibinfo {year} {2006})}\BibitemShut {NoStop}%
\bibitem [{\citenamefont {Gawlik}\ \emph {et~al.}(2006)\citenamefont {Gawlik},
  \citenamefont {Krzemień}, \citenamefont {Pustelny}, \citenamefont {Sangla},
  \citenamefont {Zachorowski}, \citenamefont {Graf}, \citenamefont {Sushkov},\
  and\ \citenamefont {Budker}}]{Gawlik2006}%
  \BibitemOpen
  \bibfield  {author} {\bibinfo {author} {\bibfnamefont {W.}~\bibnamefont
  {Gawlik}}, \bibinfo {author} {\bibfnamefont {L.}~\bibnamefont {Krzemień}},
  \bibinfo {author} {\bibfnamefont {S.}~\bibnamefont {Pustelny}}, \bibinfo
  {author} {\bibfnamefont {D.}~\bibnamefont {Sangla}}, \bibinfo {author}
  {\bibfnamefont {J.}~\bibnamefont {Zachorowski}}, \bibinfo {author}
  {\bibfnamefont {M.}~\bibnamefont {Graf}}, \bibinfo {author} {\bibfnamefont
  {A.~O.}\ \bibnamefont {Sushkov}}, \ and\ \bibinfo {author} {\bibfnamefont
  {D.}~\bibnamefont {Budker}},\ }\href {\doibase 10.1063/1.2190457} {\bibfield
  {journal} {\bibinfo  {journal} {Applied Physics Letters}\ }\textbf {\bibinfo
  {volume} {88}},\ \bibinfo {pages} {131108} (\bibinfo {year} {2006})},\
  \Eprint {http://arxiv.org/abs/https://doi.org/10.1063/1.2190457}
  {https://doi.org/10.1063/1.2190457} \BibitemShut {NoStop}%
\bibitem [{\citenamefont {Grangier}\ \emph {et~al.}(1987)\citenamefont
  {Grangier}, \citenamefont {Slusher}, \citenamefont {Yurke},\ and\
  \citenamefont {LaPorta}}]{Grangier1987}%
  \BibitemOpen
  \bibfield  {author} {\bibinfo {author} {\bibfnamefont {P.}~\bibnamefont
  {Grangier}}, \bibinfo {author} {\bibfnamefont {R.~E.}\ \bibnamefont
  {Slusher}}, \bibinfo {author} {\bibfnamefont {B.}~\bibnamefont {Yurke}}, \
  and\ \bibinfo {author} {\bibfnamefont {A.}~\bibnamefont {LaPorta}},\ }\href
  {\doibase 10.1103/PhysRevLett.59.2153} {\bibfield  {journal} {\bibinfo
  {journal} {Phys. Rev. Lett.}\ }\textbf {\bibinfo {volume} {59}},\ \bibinfo
  {pages} {2153} (\bibinfo {year} {1987})}\BibitemShut {NoStop}%
\bibitem [{\citenamefont {Smith}\ \emph {et~al.}(2008)\citenamefont {Smith},
  \citenamefont {Chu}, \citenamefont {Huang}, \citenamefont {Wiig},\ and\
  \citenamefont {Brown}}]{John2008}%
  \BibitemOpen
  \bibfield  {author} {\bibinfo {author} {\bibfnamefont {J.~A.}\ \bibnamefont
  {Smith}}, \bibinfo {author} {\bibfnamefont {X.}~\bibnamefont {Chu}}, \bibinfo
  {author} {\bibfnamefont {W.}~\bibnamefont {Huang}}, \bibinfo {author}
  {\bibfnamefont {A.~J.}\ \bibnamefont {Wiig}}, \ and\ \bibinfo {author}
  {\bibfnamefont {A.}~\bibnamefont {Brown}},\ }\href {\doibase
  10.1117/1.3013257} {\bibfield  {journal} {\bibinfo  {journal} {Optical
  Engineering}\ }\textbf {\bibinfo {volume} {47}},\ \bibinfo {pages} {47 }
  (\bibinfo {year} {2008})}\BibitemShut {NoStop}%
\bibitem [{\citenamefont {Aharonov}\ \emph {et~al.}(1988)\citenamefont
  {Aharonov}, \citenamefont {Albert},\ and\ \citenamefont
  {Vaidman}}]{Aharonov1988}%
  \BibitemOpen
  \bibfield  {author} {\bibinfo {author} {\bibfnamefont {Y.}~\bibnamefont
  {Aharonov}}, \bibinfo {author} {\bibfnamefont {D.~Z.}\ \bibnamefont
  {Albert}}, \ and\ \bibinfo {author} {\bibfnamefont {L.}~\bibnamefont
  {Vaidman}},\ }\href {\doibase 10.1103/PhysRevLett.60.1351} {\bibfield
  {journal} {\bibinfo  {journal} {Phys. Rev. Lett.}\ }\textbf {\bibinfo
  {volume} {60}},\ \bibinfo {pages} {1351} (\bibinfo {year}
  {1988})}\BibitemShut {NoStop}%
\bibitem [{\citenamefont {Duck}\ \emph {et~al.}(1989)\citenamefont {Duck},
  \citenamefont {Stevenson},\ and\ \citenamefont {Sudarshan}}]{Sudarshan1989}%
  \BibitemOpen
  \bibfield  {author} {\bibinfo {author} {\bibfnamefont {I.~M.}\ \bibnamefont
  {Duck}}, \bibinfo {author} {\bibfnamefont {P.~M.}\ \bibnamefont {Stevenson}},
  \ and\ \bibinfo {author} {\bibfnamefont {E.~C.~G.}\ \bibnamefont
  {Sudarshan}},\ }\href {\doibase 10.1103/PhysRevD.40.2112} {\bibfield
  {journal} {\bibinfo  {journal} {Phys. Rev. D}\ }\textbf {\bibinfo {volume}
  {40}},\ \bibinfo {pages} {2112} (\bibinfo {year} {1989})}\BibitemShut
  {NoStop}%
\bibitem [{\citenamefont {Ritchie}\ \emph {et~al.}(1991)\citenamefont
  {Ritchie}, \citenamefont {Story},\ and\ \citenamefont {Hulet}}]{Ritchie1991}%
  \BibitemOpen
  \bibfield  {author} {\bibinfo {author} {\bibfnamefont {N.~W.~M.}\
  \bibnamefont {Ritchie}}, \bibinfo {author} {\bibfnamefont {J.~G.}\
  \bibnamefont {Story}}, \ and\ \bibinfo {author} {\bibfnamefont {R.~G.}\
  \bibnamefont {Hulet}},\ }\href {\doibase 10.1103/PhysRevLett.66.1107}
  {\bibfield  {journal} {\bibinfo  {journal} {Phys. Rev. Lett.}\ }\textbf
  {\bibinfo {volume} {66}},\ \bibinfo {pages} {1107} (\bibinfo {year}
  {1991})}\BibitemShut {NoStop}%
\bibitem [{\citenamefont {Suna}\ and\ \citenamefont
  {Viswanathan}(2018)}]{rashmi1}%
  \BibitemOpen
  \bibfield  {author} {\bibinfo {author} {\bibfnamefont {R.~R.}\ \bibnamefont
  {Suna}}\ and\ \bibinfo {author} {\bibfnamefont {N.~K.}\ \bibnamefont
  {Viswanathan}},\ }\href {\doibase 10.1364/OL.43.004337} {\bibfield  {journal}
  {\bibinfo  {journal} {Opt. Lett.}\ }\textbf {\bibinfo {volume} {43}},\
  \bibinfo {pages} {4337} (\bibinfo {year} {2018})}\BibitemShut {NoStop}%
\bibitem [{\citenamefont {Samlan}\ \emph {et~al.}(2018)\citenamefont {Samlan},
  \citenamefont {Suna}, \citenamefont {Naik},\ and\ \citenamefont
  {Viswanathan}}]{rashmi2}%
  \BibitemOpen
  \bibfield  {author} {\bibinfo {author} {\bibfnamefont {C.~T.}\ \bibnamefont
  {Samlan}}, \bibinfo {author} {\bibfnamefont {R.~R.}\ \bibnamefont {Suna}},
  \bibinfo {author} {\bibfnamefont {D.~N.}\ \bibnamefont {Naik}}, \ and\
  \bibinfo {author} {\bibfnamefont {N.~K.}\ \bibnamefont {Viswanathan}},\
  }\href {\doibase 10.1063/1.5008732} {\bibfield  {journal} {\bibinfo
  {journal} {Applied Physics Letters}\ }\textbf {\bibinfo {volume} {112}},\
  \bibinfo {pages} {031101} (\bibinfo {year} {2018})},\ \Eprint
  {http://arxiv.org/abs/https://doi.org/10.1063/1.5008732}
  {https://doi.org/10.1063/1.5008732} \BibitemShut {NoStop}%
\bibitem [{\citenamefont {Born}\ and\ \citenamefont {Wolf}(1997)}]{Max1997}%
  \BibitemOpen
  \bibfield  {author} {\bibinfo {author} {\bibfnamefont {M.}~\bibnamefont
  {Born}}\ and\ \bibinfo {author} {\bibfnamefont {E.}~\bibnamefont {Wolf}},\
  }\href@noop {} {\emph {\bibinfo {title} {Principles of optics:
  electromagnetic theory of propagation, interference and diffraction of
  light}}}\ (\bibinfo  {publisher} {Cambridge University Press},\ \bibinfo
  {year} {1997})\BibitemShut {NoStop}%
\bibitem [{\citenamefont {Goldstein}(2010)}]{Dennis2010}%
  \BibitemOpen
  \bibfield  {author} {\bibinfo {author} {\bibfnamefont {D.~H.}\ \bibnamefont
  {Goldstein}},\ }\href@noop {} {\emph {\bibinfo {title} {Polarized Light}}},\
  \bibinfo {edition} {3rd}\ ed.\ (\bibinfo  {publisher} {CRC Press},\ \bibinfo
  {year} {2010})\BibitemShut {NoStop}%
\bibitem [{\citenamefont {Ledbetter}\ \emph {et~al.}(2008)\citenamefont
  {Ledbetter}, \citenamefont {Savukov}, \citenamefont {Acosta}, \citenamefont
  {Budker},\ and\ \citenamefont {Romalis}}]{Budker2008}%
  \BibitemOpen
  \bibfield  {author} {\bibinfo {author} {\bibfnamefont {M.~P.}\ \bibnamefont
  {Ledbetter}}, \bibinfo {author} {\bibfnamefont {I.~M.}\ \bibnamefont
  {Savukov}}, \bibinfo {author} {\bibfnamefont {V.~M.}\ \bibnamefont {Acosta}},
  \bibinfo {author} {\bibfnamefont {D.}~\bibnamefont {Budker}}, \ and\ \bibinfo
  {author} {\bibfnamefont {M.~V.}\ \bibnamefont {Romalis}},\ }\href {\doibase
  10.1103/PhysRevA.77.033408} {\bibfield  {journal} {\bibinfo  {journal} {Phys.
  Rev. A}\ }\textbf {\bibinfo {volume} {77}},\ \bibinfo {pages} {033408}
  (\bibinfo {year} {2008})}\BibitemShut {NoStop}%
\bibitem [{\citenamefont {Seltzer}\ and\ \citenamefont
  {Romalis}(2009)}]{Romalis2009}%
  \BibitemOpen
  \bibfield  {author} {\bibinfo {author} {\bibfnamefont {S.~J.}\ \bibnamefont
  {Seltzer}}\ and\ \bibinfo {author} {\bibfnamefont {M.~V.}\ \bibnamefont
  {Romalis}},\ }\href {\doibase 10.1063/1.3236649} {\bibfield  {journal}
  {\bibinfo  {journal} {Journal of Applied Physics}\ }\textbf {\bibinfo
  {volume} {106}},\ \bibinfo {pages} {114905} (\bibinfo {year} {2009})},\
  \Eprint {http://arxiv.org/abs/https://doi.org/10.1063/1.3236649}
  {https://doi.org/10.1063/1.3236649} \BibitemShut {NoStop}%
\bibitem [{\citenamefont {Knappe}\ \emph {et~al.}(2016)\citenamefont {Knappe},
  \citenamefont {Alem}, \citenamefont {Sheng},\ and\ \citenamefont
  {Kitching}}]{Sheng2016}%
  \BibitemOpen
  \bibfield  {author} {\bibinfo {author} {\bibfnamefont {S.}~\bibnamefont
  {Knappe}}, \bibinfo {author} {\bibfnamefont {O.}~\bibnamefont {Alem}},
  \bibinfo {author} {\bibfnamefont {D.}~\bibnamefont {Sheng}}, \ and\ \bibinfo
  {author} {\bibfnamefont {J.}~\bibnamefont {Kitching}},\ }\href
  {http://stacks.iop.org/1742-6596/723/i=1/a=012055} {\bibfield  {journal}
  {\bibinfo  {journal} {Journal of Physics: Conference Series}\ }\textbf
  {\bibinfo {volume} {723}},\ \bibinfo {pages} {012055} (\bibinfo {year}
  {2016})}\BibitemShut {NoStop}%
\bibitem [{\citenamefont {Sheng}\ \emph {et~al.}(2017)\citenamefont {Sheng},
  \citenamefont {Perry}, \citenamefont {Krzyzewski}, \citenamefont {Geller},
  \citenamefont {Kitching},\ and\ \citenamefont {Knappe}}]{Sheng2017}%
  \BibitemOpen
  \bibfield  {author} {\bibinfo {author} {\bibfnamefont {D.}~\bibnamefont
  {Sheng}}, \bibinfo {author} {\bibfnamefont {A.~R.}\ \bibnamefont {Perry}},
  \bibinfo {author} {\bibfnamefont {S.~P.}\ \bibnamefont {Krzyzewski}},
  \bibinfo {author} {\bibfnamefont {S.}~\bibnamefont {Geller}}, \bibinfo
  {author} {\bibfnamefont {J.}~\bibnamefont {Kitching}}, \ and\ \bibinfo
  {author} {\bibfnamefont {S.}~\bibnamefont {Knappe}},\ }\href {\doibase
  10.1063/1.4974349} {\bibfield  {journal} {\bibinfo  {journal} {Applied
  Physics Letters}\ }\textbf {\bibinfo {volume} {110}},\ \bibinfo {pages}
  {031106} (\bibinfo {year} {2017})},\ \Eprint
  {http://arxiv.org/abs/https://doi.org/10.1063/1.4974349}
  {https://doi.org/10.1063/1.4974349} \BibitemShut {NoStop}%
\end{thebibliography}%
\end{document}